# Compton Scattering Driven by Quantum Light

**Majed Khalaf and Ido Kaminer**

Department of Electrical and Computer Engineering and Solid State Institute,
Technion – Israel Institute of Technology, 32000 Haifa, Israel

**Compton scattering is one of the cornerstones of quantum physics, describing the fundamental interaction of a charged particle with photons. The Compton effect and its inverse are utilized in experiments driving free electrons by high intensity lasers to create trains of attosecond X-ray pulses. So far, all theory and experiments of the Compton effect and its generalizations have relied on electromagnetic fields that can be described classically. Advances in the generation of intense squeezed light could enable driving the Compton effect with non-classical light. This outlook motivates exploring the role of photon statistics in the Compton effect. We develop a framework to describe the full non-perturbative interaction of a charged particle with a driving field ascribed with an arbitrary quantum light state. We obtain analytical results for the Compton emission spectrum when driven by thermal and squeezed vacuum states, showing a noticeable broadening of the emission spectrum relative to a classical (coherent state) drive, thus reaching higher emission frequencies for the same average intensity. We envision utilizing the quantum properties of light, including photon statistics, squeezing, and entanglement, as novel degrees of freedom to control the wide range of radiation phenomena at the foundations of quantum electrodynamics.**

## Introduction

The scattering of light by free charged particles lies at the heart of light-matter interactions. Its most fundamental form is referred to as Compton scattering, also known as Thomson scattering in its low energy regime. Compton scattering has a wide variety of applications, ranging from clinical as in radiobiology and radiation therapy [1], to photo-nuclear reactions in nuclear physics [2] and even electron-positron pair production in areas of high energy physics [3].

When increasing the intensity of its driving field, the Compton effect transitions into its nonlinear regime, wherein multiple photons are absorbed and converted into higher energy photons. This process is called non-linear Compton scattering (NCS). NCS was studied extensively both theoretically and experimentally (see [4-6] for reviews). Theoretically, it was first studied in [7,8] for a monochromatic driving plane-wave field, and later generalized in [9, 10] for driving fields with finite spatial and temporal extensions, which model laser pulses more accurately. Experimentally, the low energy limit of NCS (also known as nonlinear Thomson scattering) was first detected via a second harmonic emission by a $1\ \mathrm{keV}$ electron beam interacting with a laser with intensity of $1.7 \cdot 10^{14}\ \mathrm{W/cm^2}$ [11]. A later experiment presented the first observation of NCS from an ultra-relativistic $46.6\ \mathrm{GeV}$ electron beam, relying on a laser intensity of $10^{18}\ \mathrm{W/cm^2}$ [12,13].

So far, all works on Compton-type effects, both theory and experiment, considered the driving laser as a classical electromagnetic field. The reason that light with non-classical photon-statistics has not been considered is the long-held conception seeing intense (many-photon) light as classical, whereas quantum states of light correspond to a small number of photons. Importantly, recent experimental advances have started to break this conception by generating light states of non-classical photon statistics that have ever-increasing intensities. Such advances motivate revisiting the Compton effects and its variants using a quantum-optical modelling of the driving field. For example, bright squeezed vacuum (BSV) picosecond pulses with an energy of $10\ \mu\mathrm{J}$ [14] and femtosecond pulses with an energy of $350\ n\mathrm{J}$ [15] have been generated through the process of parametric down conversion [16,17]. Such pulses could potentially reach intensities on the order of $10^{14}\ \mathrm{W/cm^2}$. Higher intensities could be achieved if these pulses are amplified using ultrafast amplifiers [18], which alter the quantum state but do not change the photon statistics to the Poissonian distribution of classical light as in the coherent state [16,17,19].

Looking at the big picture, even beyond Compton-type effects, all the effects of strong field physics have so far been considered without accounting for the photon statistics of light. This is the case even in astrophysical settings, where thermal radiation produced by the accretion disk of a black hole undergoes inverse Compton scattering off relativistic free electrons in the surrounding corona [18]. Even then, the effect of photon statistics was neglected (because only linear Compton scattering was considered). More generally, **the theory of non-perturbative interactions of matter with intense non-classical light is currently missing**.

Here we develop the framework for describing the non-perturbative interactions of matter with driving fields ascribed with arbitrary light states. As our primary example, we employ this framework to describe Compton scattering and the general NCS, when driven by intense light of an arbitrary quantum state (Fig. 1). Though the framework applies to any fermionic charged particle, we focus on the example of Compton scattering off free electrons. We calculate the NCS emission spectrum and angular distribution, identifying the unique aspects of non-classical driving light states. Specifically, we obtain analytical formulas for driving by intense thermal and BSV states. We find that the resulting spectra are broadened relative to the usual spectrum of classically (coherent-state) driven NCS of the same intensity, reaching far higher frequencies in some regimes of parameters. Moreover, in sharp deviation from the conventional Compton scattering, the spectrum per solid angle can become continuous.

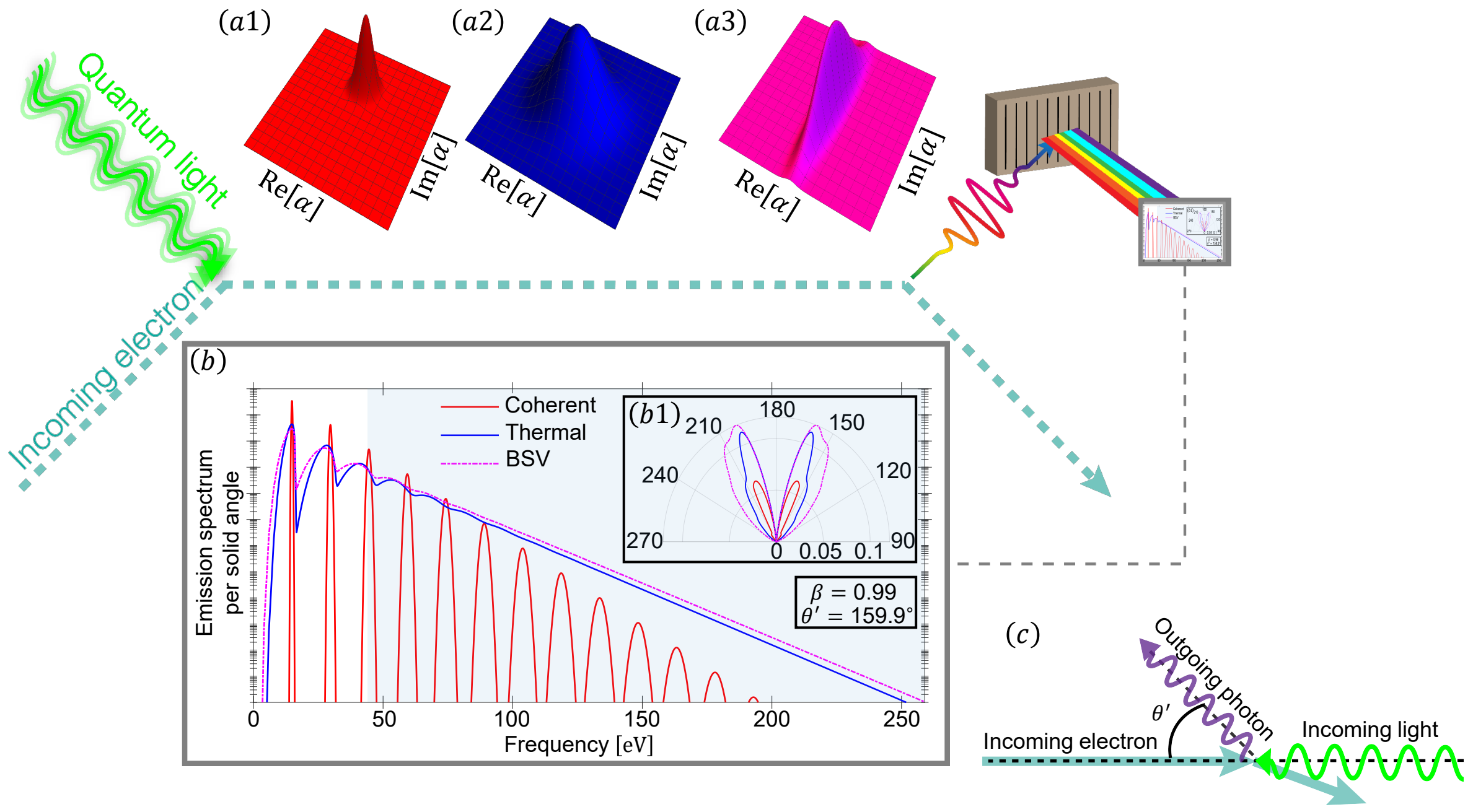

**Fig. 1 | Non-linear Compton scattering driven by light with quantum photon statistics: spectrum broadening.** Consider a free electron that scatters off different intense light states described by the Husimi functions in (a1-a3). We consider the cases of a coherent state (a1, red color), a thermal state (a2, blue color), and a bright squeezed vacuum (BSV) state (a3, magenta color). The resulting NCS spectrum per solid angle and angular distribution appear in (b) and (b1), respectively. Panel (c) defines the notation for the polar angle $\theta'$ of photon emission. The angular distribution of emission in (b1) was obtained after integrating over the frequency range that falls in the blue shaded region marked in (b). The spectrum of (b) is plotted at $\theta' = 159.9°$, chosen at the peak of emission in (b1). $\beta = 0.99$ is the velocity of the electron normalized by the velocity of light. Altogether, panel (b) shows how manipulating photon statistics can result in a noticeable spectrum broadening and its extension to higher frequency radiation. In the examples we plot, the incoming electron is counter-propagating to the incoming light. All the results in the main text are general to any incoming angle and any combination of parameters.

**Non-perturbative interaction of matter with arbitrary light states in QED**

In this section, we introduce the framework for the non-perturbative treatment of the interaction of matter with arbitrary quantum states of light. Unlike the dynamics of the strong driving light field, the spontaneous emission by the joint state of matter and the driving field can be treated perturbatively. We show that the result can be constructed from solving for the spontaneous emission by the same matter state with a range of external classical light states (coherent states).

Let $\mathcal{H} = \mathcal{H}_{\text{matter}} + \mathcal{H}_{\text{int}} + \mathcal{H}_{\text{F}}$ be the Hamiltonian of the system. Here, $\mathcal{H}_{\text{matter}}$ is the matter Hamiltonian, $\mathcal{H}_{\text{int}}$ is the interaction Hamiltonian between matter and the quantized electromagnetic field, and $\mathcal{H}_{\text{F}} = \sum_{\vec{k}',\epsilon'} \hbar\omega' a^{\dagger}_{\vec{k}',\epsilon'} a_{\vec{k}',\epsilon'}$ is the field Hamiltonian, where the summation is over the wavevectors $\vec{k}'$ and polarizations $\epsilon'$. The wavevectors satisfy the free-space dispersion relation with the photon frequency $\omega' = c|\vec{k}'|$. The operators $a_{\vec{k}',\epsilon'}$, $a^{\dagger}_{\vec{k}',\epsilon'}$ are the photonic annihilation and creation operators, satisfying the commutation relation $\left[a_{\vec{k}',\epsilon'}, a^{\dagger}_{\vec{k}',\epsilon'}\right] = 1$.

The only assumption required for our general solution below is having $\mathcal{H}_{\text{int}}$ linear in $a_{\vec{k}',\epsilon'}$ and $a^{\dagger}_{\vec{k}',\epsilon'}$, as is the case in conventional light-matter interactions and in all the fundamental processes of quantum electrodynamics like the Compton effect. The field energy $\mathcal{H}_{\text{F}}$ will be henceforth omitted from the Hamiltonian because we choose to transform it away using $\exp[-it\mathcal{H}_{\text{F}}/\hbar]$, at the expense of making the operators $a_{\vec{k}',\epsilon'}$ and $a^{\dagger}_{\vec{k}',\epsilon'}$ in $\mathcal{H}_{\text{int}}$ explicitly time-dependent.

Let $\rho_{\text{light}}$ denote the initial density matrix of the incoming light. Previous experiments of NCS showed that its main features can be well-approximated by a single mode (plane wave), which we denote here by wavevector $\vec{k}$ and polarization $\epsilon$. We can express $\rho_{\text{light}}$ using the generalized Glauber distribution $P(\alpha,\beta)$ [21] as follows:

$$\rho_{\text{light}} = \int d^2\alpha \, d^2\beta \, P(\alpha,\beta) \frac{|\alpha\rangle_{\vec{k},\epsilon}\langle\beta^*|_{\vec{k},\epsilon}}{\langle\beta^*|\alpha\rangle_{\vec{k},\epsilon}} \,. \tag{1}$$

where $|\alpha\rangle_{\vec{k},\epsilon}$ is a coherent state of parameter $\alpha$ in the $\vec{k},\epsilon$ mode. We note that $P(\alpha,\beta)$ can be chosen to be regular and positive [21].

Using Eq. (1) and the linearity of the time evolution equation, we find that the density matrix of the combined light-matter system at time $t$ can be written as

$$\rho_{\text{total}}(t) = \int d^2\alpha\, d^2\beta\, \frac{P(\alpha,\beta)}{\langle\beta^*|\alpha\rangle_{\vec{k},\epsilon}} D_{\vec{k},\epsilon}(\alpha)|\Psi^\alpha(t)\rangle\langle\Psi^{\beta^*}(t)|\, D^\dagger_{\vec{k},\epsilon}(\beta^*)\,. \tag{2}$$

$D_{\vec{k},\sigma}$ is the displacement operator [16,17] that acts only on the $\vec{k},\epsilon$ mode, and $|\Psi^\alpha(t)\rangle$ is the state of the system at time $t$ with an initial condition $|\Psi^\alpha(-\infty)\rangle = |\varphi_{\text{i}}\rangle \otimes |0\rangle$, made from an initial matter state $|\varphi_{\text{i}}\rangle$ and no photons. The dynamics of $|\Psi^\alpha(t)\rangle$ is dictated by the time-dependent Schrodinger equation with the Hamiltonian $\widetilde{\mathcal{H}}^\alpha = \mathcal{H}_{\text{matter}} + \mathcal{H}^\alpha_{\text{int}} + \mathcal{H}^{\text{SE}}_{\text{int}}$. A time-dependent semiclassical interaction Hamiltonian $\mathcal{H}^\alpha_{\text{int}}$ is obtained from $\mathcal{H}_{\text{int}}$ by replacing $a_{\vec{k},\epsilon}$ with $\alpha$ and discarding the other modes, leaving a small remainder $\mathcal{H}^{\text{SE}}_{\text{int}}$ that is responsible for the spontaneous emission by the time-dependent Hamiltonian. In other words, $|\Psi^\alpha(t)\rangle$ evolves in time as though the matter is not only interacting with the quantized field, but also with an external classical field represented by $\alpha$.

No approximations were done to derive Eq. (2). However, it is too general to be useful for our purposes here. To cast it into a more useful form, we henceforth treat the interaction with $\mathcal{H}^{\text{SE}}_{\text{int}}$ perturbatively. Using first order perturbation theory yields

$$|\Psi^\alpha(t)\rangle = |\varphi^\alpha_{\text{i}}(t)\rangle \otimes |0\rangle + \sum_{f,\vec{k}',\epsilon'} S^\alpha_{\text{i}\to f,\vec{k}',\epsilon'} \left|\varphi^\alpha_f(t)\right\rangle \otimes \left|1_{\vec{k}',\epsilon'}\right\rangle. \tag{3}$$

Here, $\left|1_{\vec{k}',\epsilon'}\right\rangle$ is a single photon Fock state in the $\vec{k}',\epsilon'$ mode and $S^\alpha_{\text{i}\to f,\vec{k}',\epsilon'}$ is the amplitude for the matter to transition from the state $\varphi^\alpha_{\text{i}}$ to any possible final state $\varphi^\alpha_f$, accompanied by the emission of a photon with wavevector $\vec{k}'$ and polarization $\epsilon'$. The index $f$ enumerates over all matter states $\{\varphi^\alpha_f\}_f$, which are orthogonal non-stationary states satisfying

$$i\hbar\frac{\partial}{\partial t}\left|\varphi^\alpha_f(t)\right\rangle = (\mathcal{H}_{\text{matter}} + \mathcal{H}^\alpha_{\text{int}})\left|\varphi^\alpha_f(t)\right\rangle, \tag{4}$$

with the initial condition that each $\left|\varphi^\alpha_f(-\infty)\right\rangle$ is an eigenstate of $\mathcal{H}_{\text{matter}}$. We note that Eq. (3) shows how the interaction process creates entanglement between the light and matter parts of the joint quantum state. Thus, to fully understand the light-matter entangled state, it is necessary to calculate $S^\alpha_{\text{i}\to f,\vec{k}',\epsilon'}$. For the case of Compton scattering (including NCS), these $S$ factors are presented in Supplementary Materials (SM) section 2.1.

Eqs. (3-4) convert the general QED problem into a semiclassical time-dependent Schrodinger equation that can be solved numerically. While this framework can be used to calculate any observable, we exemplify it below by presenting the emission spectrum. Since first order perturbation theory is the leading contribution to the spectrum, we can use Eqs. (2-3) directly to find (see SM section 1 for details)

$$\frac{d\varepsilon}{d\omega' d\Omega'} = \frac{\hbar\omega'^3}{(2\pi c)^3} \sum_{f,\epsilon'} \int d^2\mathcal{E}_\alpha \, \tilde{Q}(\mathcal{E}_\alpha) \left| \tilde{S}^\alpha_{\mathrm{i}\to f,\vec{k}',\epsilon'} \right|^2 . \tag{5}$$

Here, $\Omega'$ is the emitted photon solid angle, $\tilde{S}^\alpha_{\mathrm{i}\to f,\vec{k}',\epsilon'} = \sqrt{V} S^\alpha_{\mathrm{i}\to f,\vec{k}',\epsilon'}$ where $V$ is the photonic quantization volume, and $\mathcal{E}_\alpha = \alpha\sqrt{2\hbar\omega/\epsilon_0 V}$ is the electric field amplitude of the state $|\alpha\rangle_{\vec{k},\epsilon}$. We denote by $\tilde{Q}(\mathcal{E}_\alpha)$ the limit of $\frac{V\epsilon_0}{2\hbar\omega} Q\left(\sqrt{\frac{V\epsilon_0}{2\hbar\omega}}\mathcal{E}_\alpha\right)$ as $V \to \infty$, such that $\mathcal{E}_\alpha$ and the average photon number density of the driving field are kept constant, with $Q$ being the Husimi representation of $\rho_{\mathrm{light}}$. We note that $\tilde{S}^\alpha_{\mathrm{i}\to f,\vec{k}',\epsilon'}$ is independent of $V$, so that Eq. (5) is independent of $V$.

Throughout the remainder of this manuscript, we focus on the example of Compton scattering. Hence, we apply the formalism discussed above in the case where the matter is comprised of a single charged particle (henceforth shown for electrons), with the free Dirac Hamiltonian [22] acting as $\mathcal{H}_{\mathrm{matter}}$. An electron moving in a periodic potential emits as long as the motion continues, and the spectrum of the total emitted radiation energy increases linearly in time. We therefore consider the spectrum of emitted power rather than energy, obtained by dividing the spectrum of the total emitted radiation energy Eq. (5) by the total duration of motion $T$. Furthermore, we use the natural unit system where Planck's constant $\hbar$, velocity of light in vacuum $c$, and the vacuum permittivity $\varepsilon_0$ are set to 1.

**The emission power spectrum of Compton scattering driven by arbitrary light states**

In this section, we derive formulas for the emission power spectrum of Compton scattering driven by arbitrary light states. We show that the photon statistics can drastically change the spectrum. In particular, we reproduce the conventional Compton scattering spectrum (driven by classical light), and calculate the spectrum for thermal- and BSV- driven Compton Scattering. Our only assumption is for the polarization of the driving field to be circular $\epsilon^\mu = \frac{1}{\sqrt{2}}(0,1,i,0)$, since it simplifies the analytical expressions (all results can be directly extended to any polarization).

The emitted photon modes are described by a null four-wavevector $k^{\mu\prime} = (\omega', \vec{k}') = \omega'(1, \sin\theta' \cos\phi', \sin\theta' \sin\phi', \cos\theta')$ and a unit complex polarization four-vector $\epsilon'^{\mu}$. $k^\mu = (\omega, \vec{k})$ is the four-wavevector of the driving field and $\epsilon^\mu$ is its polarization. We employ Eq. (5) to derive a formula for the emission power spectrum (see SM section 2). We assume that the electron's spin is not measured so we average over the initial spin of the electron. The result turns out to be

$$\frac{1}{T}\frac{d\varepsilon}{d\omega' d\Omega'} = \frac{\omega^2 \omega'^2}{4\pi^2}\frac{k\cdot p}{e^2 k\cdot k'}\, p^{t'} \sum_{s=1}^{\infty} \theta(s(p\cdot k - k\cdot k') - p\cdot k')\, \langle |T_s|^2 \rangle \int_0^{2\pi} d\phi\, \tilde{Q}(\mathcal{E}_s e^{i\phi}) \;, (6)$$

where $-e$ is the electron's charge, $s$ is the emission order, $m_e$ is the mass of the electron, $p^\mu$ is the initial four-momentum of the electron, $J_l$ is the Bessel function of the first kind of order $l$, and $p^t$ and $p^{t'}$ are the timelike components ($\mu = 0$ or $t$) of $p^\mu$ and $p^{\mu'}$, respectively. The outgoing electron four-momentum is derived by conservation of energy and momentum $p^{\mu'} = p^\mu + \frac{p\cdot k'}{p\cdot k - k\cdot k'} k^\mu - k^{\mu'}$ and depends on the outgoing photon four-momentum $k^{\mu'}$.

The parameters $\zeta_s$, $\xi_s$, and $T_s$ are defined in the following way

$$\begin{cases} \langle |T_s|^2 \rangle = \dfrac{e^2 m_e^2}{p^t p^{t'}}\left[\zeta_s \dfrac{(p'\cdot k)^2 + (p\cdot k)^2}{2m_e^2 k\cdot k'}\left(J_{s-1}^2(\xi_s) + J_{s+1}^2(\xi_s) - 2J_s^2(\xi_s)\right) - J_s^2(\xi_s)\right], \\ \zeta_s = \dfrac{e^2 \mathcal{E}_s^2}{4\omega^2}\left(\dfrac{1}{k\cdot p'} - \dfrac{1}{k\cdot p}\right), \\ \xi_s = e\dfrac{\mathcal{E}_s}{\omega}\left|\dfrac{p\cdot\epsilon}{k\cdot p} - \dfrac{p'\cdot\epsilon}{k\cdot p'}\right|. \end{cases} \quad (7)$$

They all depend on the parameter $\mathcal{E}_s$ that can be thought of as an effective electric field amplitude. The emission power is determined by the driving field photon statistics $\tilde{Q}$ at this effective field $\mathcal{E}_s$. This effective electric field amplitude is itself determined by the scattering problem, and depends on the emission order $s$ and on the four-momenta of the participating particles $\mathcal{E}_s^2 = \frac{4\omega^2 k\cdot p}{e^2 k\cdot k'}\,(s(p\cdot k - k\cdot k') - p\cdot k')$. The dot product symbol denotes $a\cdot b = \eta_{\mu\nu} a^\mu b^\nu$, where our convention for the Minkowski metric is $\eta_{\mu\nu} = \mathrm{diag}(1, -1, -1, -1)$.

It is not hard to see (SM section 1.2) that $\int_0^{2\pi} d\phi\, \tilde{Q}(\mathcal{E}_s e^{i\phi})$ depends only on the photon statistics of $\rho_{\text{light}}$. Thus, looking at Eq. (6), we see that it is the photon statistics, rather than the full state, that determine the emission spectrum. Unlike the spectrum, higher order correlations in the emission depend on additional properties of the driving light state besides its photon statistics, such as higher order coherence and off-diagonal elements of its density matrix.

We can use Eq. (6) to find the power spectrum for any driving light state. We analyze the cases of coherent, thermal, and BSV states as examples. To compare them, we take them to have the same average photon number density $\rho_{\text{photons}}$ (equivalent to having the same intensity). For each case, we substitute its relevant $\tilde{Q}(E)$ in Eq. (6) to obtain the power spectrum ($E$ is a complex variable with electric field units). The results are summarized in Table 1 (see SM section 2 for detailed calculations). We note that the power spectrum in the coherent case reproduces the conventional NCS result [7,23], as expected. The results in the thermal and BSV cases are found here for the first time, to the best of our knowledge.

| | |
|---|---|
| **Coherent light** | |
| $\tilde{Q}(E) = \delta^2\left(E - \sqrt{2\omega\rho_{\text{photons}}}\right)$ | $(8a)$ |
| $\frac{1}{T}\frac{d\varepsilon}{d\omega' d\Omega'} = \frac{\omega\omega'^2 p^{t'}}{4\pi^2\rho_{\text{photons}}}\frac{k\cdot p}{e^2 k\cdot k'}\sum_{s=1}^{\infty}\delta\left(\frac{\mathcal{E}_s^2}{2\omega\rho_{\text{photons}}}-1\right)\langle\lvert T_s\rvert^2\rangle$ | $(8b)$ |
| **Thermal light** | |
| $\tilde{Q}(E) = \frac{1}{2\pi\omega\rho_{\text{photons}}}\exp\left(-\frac{\lvert E\rvert^2}{2\omega\rho_{\text{photons}}}\right)$ | $(9a)$ |
| $\frac{1}{T}\frac{d\varepsilon}{d\omega' d\Omega'} = \frac{\omega\omega'^2 p^{t'}}{4\pi^2\rho_{\text{photons}}}\frac{k\cdot p}{e^2 k\cdot k'}\sum_{s=1}^{\infty}\exp\left(-\frac{\mathcal{E}_s^2}{2\omega\rho_{\text{photons}}}\right)\theta(s(p\cdot k - k\cdot k') - p\cdot k')\,\langle\lvert T_s\rvert^2\rangle$ | $(9b)$ |
| **BSV light** | |
| $\tilde{Q}(E) = \frac{1}{2\lvert E\rvert\sqrt{\pi\omega\rho_{\text{photons}}}}\delta(\cos\sphericalangle E)\exp\left(-\frac{\lvert E\rvert^2}{4\omega\rho_{\text{photons}}}\right)$ | $(10a)$ |
| $\frac{1}{T}\frac{d\varepsilon}{d\omega' d\Omega'} = \frac{\omega^{3/2}\omega'^2 p^{t'}}{4\pi^2\sqrt{\pi\rho_{\text{photons}}}}\frac{k\cdot p}{e^2 k\cdot k'}\sum_{s=1}^{\infty}\exp\left(-\frac{\mathcal{E}_s^2}{4\omega\rho_{\text{photons}}}\right)\theta(s(p\cdot k - k\cdot k') - p\cdot k')\frac{\langle\lvert T_s\rvert^2\rangle}{\mathcal{E}_s}$ | $(10b)$ |

**Table. 1 | Power spectrum formulas for coherent-, thermal-, and BSV-driven Compton scattering.**

Eqs. (8b-10b) show that photon statistics can significantly change the resulting spectrum formula. The most striking feature, perhaps, is that unlike conventional Compton scattering and NCS that have discrete peaks in the spectrum for each solid angle (Eq. (8b)), the thermal- and BSV-driven Compton effects have continuous spectra for each solid angle (Eqs. (9b-10b)).

## Spectral broadening and higher harmonics induced by photon statistics

In this section, we compare between Eqs. (8b-10b) and show that the power spectrum is broadened and reaches higher frequencies for the thermal and BSV cases, compared with the coherent-state case of the same intensity. i.e., we conclude that driving the Compton effect with non-classical photon statistics can enable reaching higher emission frequencies for the same average intensity. We consider two distinctive regimes: Compton scattering, in which the initial electron is at rest, and inverse-Compton scattering, in which the initial electron is counter-propagating to the driving pulse at a relativistic velocity, up-converting the radiation frequency. In Compton scattering, the driving field transfers part of its energy to the electron. In inverse-Compton scattering, the energetic electron transfers part of its energy to the emitted photon.

Eqs. (8b-10b) were obtained assuming a monochromatic driving field, which approximates realistic experiments but requires corrections arising from the pulsed (and thus spectrally wide) nature of light usually needed to achieve high intensity fields. To model a more realistic scenario of a field centered around the frequency $\omega$ with width $\Delta_\omega$ (assumed $\ll \omega$), we convolve Eqs. (8b-10b) with a Gaussian profile with variance $\Delta_\omega^2$. Then, to get the energy spectrum, we multiply the power spectrum by the driving pulse's characteristic time interval, assuming a Fourier-limited pulse duration $T = 2\pi/\Delta_\omega$.

For the remainder of this section, we take $\omega = 2.25$ eV (green light) and $\Delta_\omega/\omega = 8 \cdot 10^{-3}$ (corresponds to pulse duration of 229.75 fs). The propagation direction of the driving field is in the $z$ direction. We calculate the energy spectrum and the energy angular distribution from Eqs. (8b-10b) for Compton and inverse-Compton scattering, presented in Figs. 2 and 3, respectively.

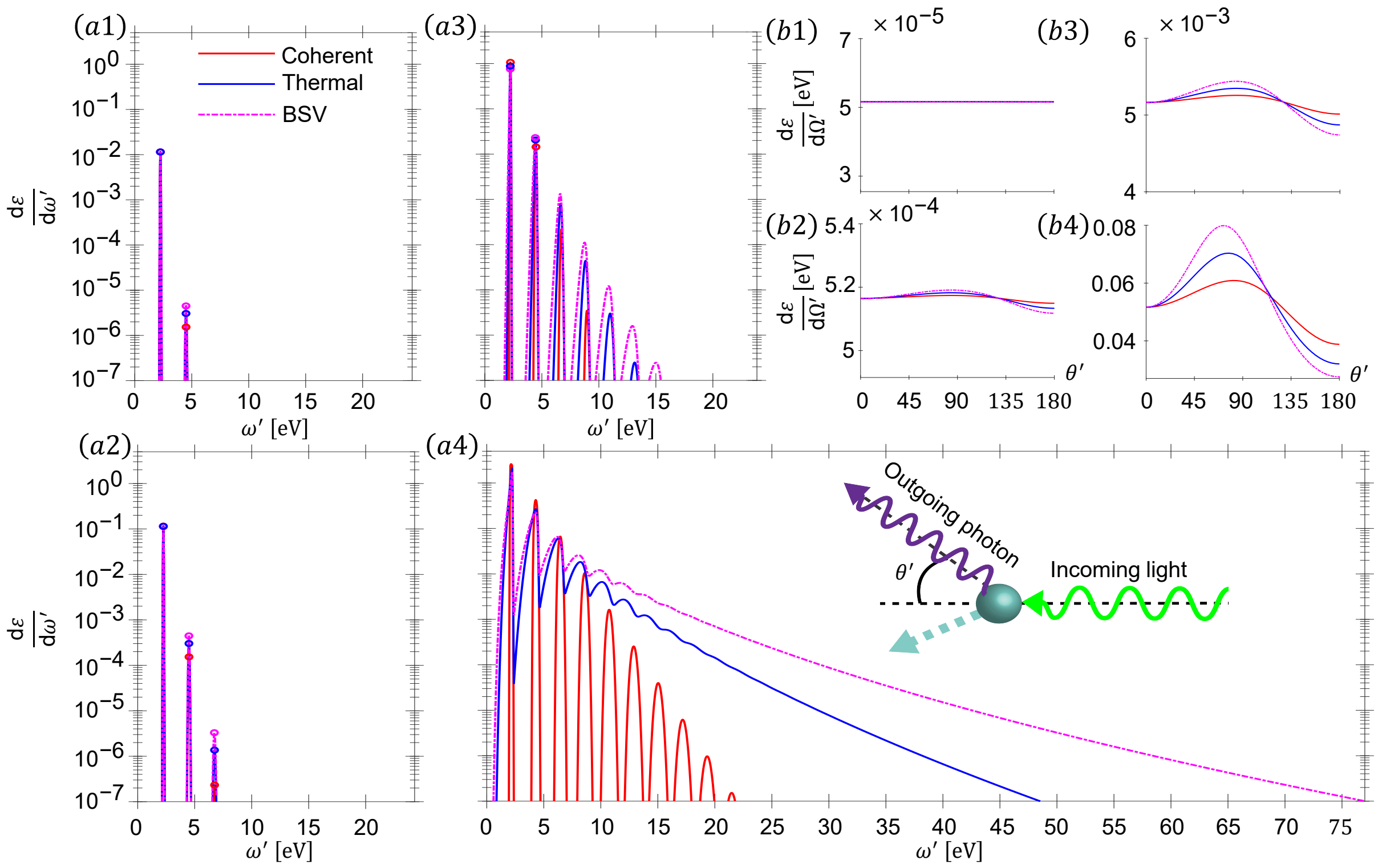

**Fig. 2 | The effect of photon statistics on the spectrum and angular distribution of Compton scattering.** We assume a stationary electron interacting with different driving field intensities. The solid red curves stand for the coherent state, solid blue for the thermal state, and dashed magenta for the bright squeezed vacuum (BSV) state. Panels (a1-a4) show the energy spectra of Compton scattering for different driving field intensities: $9 \cdot 10^{14}$ W/cm$^2$, $9 \cdot 10^{15}$ W/cm$^2$, $9 \cdot 10^{16}$ W/cm$^2$, and $9 \cdot 10^{17}$ W/cm$^2$, respectively. Panels (b1-b4) show the corresponding angular distributions as a function of the polar angle $\theta'$ defined schematically in the inset (dashed blue arrow represents the outgoing electron).

Panels (a1-a4) of Fig. 2 show how the energy spectrum in Compton scattering changes when increasing the driving intensity (tenfold between sequential panels). We see that BSV- and thermal-driven Compton scattering outperform the coherent-driven one in terms of the spectrum width: both the width of individual spectral peaks, and the overall extent of the spectrum. i.e., BSV-driven and thermal-driven Compton scattering reach higher harmonics of the driving field frequency than conventional Compton scattering driven by the same intensity. This spectral broadening is most notably seen in Fig. 2d, where for the BSV case the overall spectrum extends more than three times farther than for the coherent case. For the thermal case, the spectrum extends roughly two times farther than for the coherent case. The corresponding angular distributions in panels (b1-b4) of Fig. 2 show that the emission becomes more directional for higher driving intensities, being the largest for the BSV case, followed by the thermal case, and then the coherent case.

Panels (a-d) of Fig. 3 show similar phenomena for inverse-Compton scattering, represented by a relativistic electron with a Lorentz factor $\gamma \approx 7.09$ counter-propagating to the driving light pulse. The results are presented for the same driving field intensities as in Fig. 2. Similar to Compton scattering, the spectrum extends farther, reaching higher frequencies in the BSV-driven and thermal-driven cases than in conventional inverse-Compton scattering. This spectral enhancement is especially large at higher frequencies, as seen in panel (c1) that presents the corresponding angular distribution of panel (c), integrating over the frequencies marked by the blue shaded region. The angular distribution shows an order of magnitude increase of emission energy for BSV- and thermal-driven inverse Compton relative to conventional inverse-Compton driven by the same intensity. This extension of the spectrum and emission energy enhancement seem to come at the price of reducing the emission energy at lower frequencies, as seen in panel (d1) that presents the corresponding angular distribution of panel (d) integrated over the lower frequencies marked by the blue shaded region.

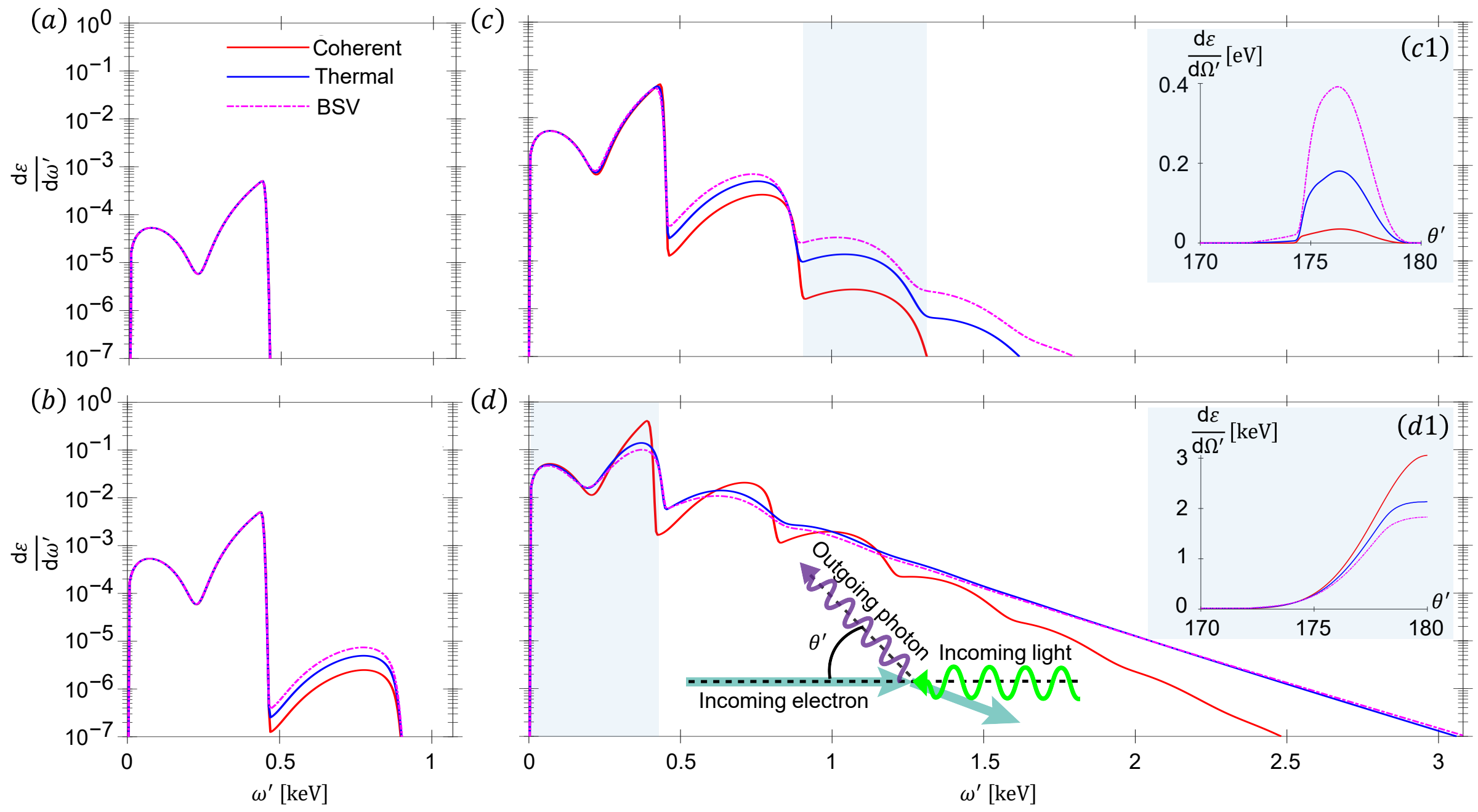

**Fig. 3| The effect of photon statistics on the spectrum and angular distribution of inverse Compton scattering.** We assume a relativistic electron (3.6 MeV, i.e., Lorentz factor $\gamma \approx 7.09$) interacting with different driving field intensities. The solid red curves stand for the coherent state, solid blue for the thermal state, and dashed magenta for the bright squeezed vacuum (BSV) state. Panels (a-d) show the energy spectra of inverse-Compton scattering for different driving field intensities: $9 \cdot 10^{14}\ \mathrm{W/cm^2}$, $9 \cdot 10^{15}\ \mathrm{W/cm^2}$, $9 \cdot 10^{16}\ \mathrm{W/cm^2}$, and $9 \cdot 10^{17}\ \mathrm{W/cm^2}$, respectively. Panels (c1) and (d1) show corresponding angular distributions as a function of the polar angle $\theta'$ defined schematically in the inset (dashed blue arrow represents the outgoing electron). These angular distributions were obtained by integrating Eqs. (7b-9b) over the frequency ranges marked by the blue shaded regions in (c) and (d).

## Discussion

The predictions presented in this work are within reach of current experimental capabilities. The main challenge is producing intense enough pulses of thermal or BSV light. An intensity of the order $\sim 10^{14}\ \mathrm{W/cm^2}$ is enough to produce a second harmonic ($s = 2$) in NCS [11]. Hence, an intensity on the order of $10^{16}\ \mathrm{W/cm^2}$ should be enough to observe the spectrum broadening seen in Figs. 2(a3,a4) and 3(c,d). While thermal or BSV pulses of such intensities have yet to be produced, BSV pulses of intensities on the order of $10^{14}\ \mathrm{W/cm^2}$ are already producible through the process of parametric down conversion [14,15]. These pulses can potentially be amplified further using ultrafast amplifiers [18]. Such an amplification stage alters the photon statistics but does not convert it to be Poissonian, and thus the features we predicted could be observed. Another approach to access the predictions of our work could be in astrophysical settings such as accretion disks of black holes [24], which could contain thermal light of extreme intensities.

Eq. (6) shows that for two driving light states to have the same Compton scattering spectrum, they must have equal values of $\int_0^{2\pi} d\phi\ \tilde{Q}(\mathcal{E}_s e^{i\phi})$, which can occur even if they have different photon statistics. For example, the state $\rho_{\text{light}} = \frac{1}{\sqrt{2\langle n\rangle+1}} \sum_{n=0}^{\infty} \left(\frac{2\langle n\rangle}{2\langle n\rangle+1}\right)^n \frac{(2n)!}{4^n (n!)^2} |n\rangle\langle n|$ has the Husimi function $Q(\alpha) = \frac{1}{\pi\sqrt{2\langle n\rangle+1}} \exp\left(-\frac{\langle n\rangle+1}{2\langle n\rangle+1}|\alpha|^2\right) I_0\left(\frac{\langle n\rangle|\alpha|^2}{2\langle n\rangle+1}\right)$, which has the same $\tilde{Q}$ as the BSV case, and consequently its spectrum is given by Eq. (10b). Here, $\langle n\rangle$ is the average photon number of the state (we identify $\rho_{\text{photons}} = \langle n\rangle/V$), and $I_0$ is the modified Bessel function of the first kind of order 0. Another example is the large $n$ Fock state $|n\rangle$, which has the same $\int_0^{2\pi} d\phi\ \tilde{Q}(\mathcal{E}_s e^{i\phi})$ as the coherent state (SM section 2.6), and thus its spectrum is equivalent to conventional Compton (Eq. (8b)). Finally, the Schrodinger cat states also have the same $\int_0^{2\pi} d\phi\ \tilde{Q}(\mathcal{E}_s e^{i\phi})$ as the coherent state (section 2.5), and thus they have the same Compton spectrum.

The consequence of having the entire emission spectrum decided solely by the photon statistics may seem at first glance to trivialize the result. An experiment can mimic the same emission spectra of Eqs. (9b-10b) using an incoherent average over classical light pulses of different intensities, for example using an attenuator that will create the same classical distribution of intensities as the diagonal of the density matrices of the thermal or BSV cases. However, this approach will require far higher intensities to start with before attenuation, and higher intensities in part of the pulses after attenuation.

An especially promising avenue for an experimental demonstration of our predicted spectral broadening and higher harmonics is using intense thermal light pulses. While this may seem like a simple task, to the best of our knowledge there are no current laboratory experiments that create thermal-light pulses of sufficient intensities for driving NCS. We propose using high intensity amplifiers seeded by short BSV pulses, creating approximations of thermal light (with certain variations of the thermal photon statistics) that will still achieve the spectral broadening and higher harmonics. We find it intriguing that despite thermal light being generally considered as completely classical, and usually not expected to provide any added value beyond coherent states, it can now enable new kinds of experiments and applications. It would be interesting to consider further applications of thermal-like photon statistics for driving phenomena of strong-field nonlinear optics and non-perturbative quantum electrodynamics.

**Conclusions and outlook**

In this work, we showed that photon statistics can noticeably change the features of the Compton emission spectrum. We obtained analytical results for thermal- and BSV-driven Compton scattering, including its complete higher-order nonlinear regime known as NCS. Comparing Eqs. (9b-10b) with conventional Compton scattering (Eq. (8b)), we notice two striking differences: (1) the emission spectrum reaches higher harmonics, and (2) the spectrum for each emission angle becomes continuous rather than discrete as in the case of a classical-light drive.

Figs. 2 and 3 show the change in the features of the emission spectrum for Compton scattering and inverse-Compton scattering, respectively. The considered simulation parameters do not exhaust all the possible regimes of Compton-type effects, and other combinations of parameters can still lead to exciting phenomena that depend on the photon statistics, especially in unique regimes of NCS such as its "quantum regime" [5,23] (there, the quantumness is not in the sense of quantum optics, but in the sense of $\hbar$-dependent corrections to the dispersion relation at high energies). Eqs. (8b-10b) can be directly used in these regimes. These quantum effects can now be combined with features that arise from the quantum-optical nature of light, such as its photon statistics.

Although this work focused primarily on the Compton spectrum, the framework described here is more general. For example, it can be used to create higher order photonic correlations in the emitted X-ray or gamma-rays, or to induce electron-photon entanglement as found in Eq. (3).

Looking at the bigger picture, the ideas presented here are not limited to Compton scattering, and can be carried over to other platforms involving intense driving light fields, such as high harmonic generation [25] and above-threshold ionization [26]. Essentially, the entirety of the field of nonlinear optics – developed over the 20th century – can now be revisited when driven by non-classical light. Even more than nonlinear optics, a wide range of fundamental effects in physics such as Rabi oscillations and the photoelectric effect can now be revisited when driven by light with non-classical photon statistics. These novel possibilities hint at a new research field to be explored, at the borderline between quantum optics and strong-field nonlinear optics.